\documentclass[aps,prd,superscriptaddress,showpacs,preprint]{revtex4}
\usepackage{graphicx, bm}
\usepackage{axodraw}
\usepackage{epsfig}
\usepackage{amsbsy}
\usepackage{varioref}
\usepackage{pifont}
\usepackage{axodraw}
\usepackage{epsfig}
\usepackage{amsbsy}
\usepackage{varioref}
\usepackage{pifont}
\usepackage{booktabs}
\usepackage{color}

\newcommand{\hs}{\hspace*{0.5cm}}

\newcommand{\be}{\begin{equation}}
\newcommand{\ee}{\end{equation}}
\newcommand{\bea}{\begin{eqnarray}}
\newcommand{\eea}{\end{eqnarray}}
\newcommand{\nn}{\nonumber}
\newcommand{\crn}{\nonumber \\}

\newcommand{\al}{\alpha}
\newcommand{\la}{\lambda}

\newcommand{\ga}{\gamma}

\newcommand{\om}{\omega}

\newcommand{\fr}{\frac}

\newcommand{\bc}{\begin{center}}
\newcommand{\ec}{\end{center}}

\newcommand{\ep}{\epsilon}

\newcommand{\ph}{\phi}

\newcommand {\ba}{\begin{array}}
\newcommand {\ea}{\end{array}}
\newcommand{\ben}{\begin{enumerate}}
\newcommand{\een}{\end{enumerate}}

\begin{document}
\draft
\title{Bounds on  dipole moments of  tau-neutrino from single photon searches in  $SU(4)_L\times U(1)_X$ model at  CLIC and ILC energies }
\author{ D. T. Binh}
\email{dinhthanhbinh@tdt.edu.vn}
\affiliation{\small Theoretical Particle Physics and Cosmology Research Group, Advanced Institute of Materials Science,   Ton Duc Thang University, Ho Chi Minh City, Vietnam\\}
\affiliation{Faculty of Applied Sciences, Ton Duc Thang University, Ho Chi Minh City, Vietnam}
\author{Vo Van On}
\email{onvv@tdmu.edu.vn}
\affiliation{Department of Physics, Faculty of Natural Sciences,
University of Thu Dau Mot, Binh Duong, Vietnam}
\affiliation{Group of Computational Physics, Faculty of Natural Sciences,  University of Thu Dau Mot, Binh Duong, Vietnam}
\author{H. N. Long}
\email{hnlong@iop.vast.ac.vn}
\affiliation{Institute of Physics, Vietnam Academy of Science and Technology,  10 Dao Tan, Ba Dinh,  Hanoi, Vietnam}
\affiliation{Department of Physics, Faculty of Natural Sciences,
University of Thu Dau Mot, Binh Duong, Vietnam}
\date{\today}

\begin{abstract}

We investigate the dipole moments of the tau- neutrino at high-energy and high luminosity at linear electron positron colliders, such as CLIC or ILC through the analysis of the  reaction
$e^{+}\, e^{-}\rightarrow \nu\, \bar \nu \,\gamma$ in the framework
of the $SU(4)_L \times U(1)_X$ model. The limits on dipole moment were obtained for integrated luminosity of $\cal{L}$= 500-2000\,  fb$^{-1}$ and center of mass between 0.5 and 3.0 TeV. The estimated limits for the tau-neutrino magnetic and
electric dipole moments  are improved by 2-3 orders of magnitude
and complement previous studies on the dipole moments.
\end{abstract}


\maketitle

\section{Introduction}

Experiment and theoretical studies of neutrino oscilation in solar \cite{SNO}, atmospheric \cite{ATM}  gave strong evidence of the non-zero mass of neutrino.  A massive neutrino can have non-trivial electromagnetic properties through radiative correction, and  If so
  a neutrino coupling to photons becomes possible. The most  important of electromagnetic processes of the direct neutrino couplings with photons are
\begin{itemize}
\item The radiative decay of a heavier neutrino
into a lighter neutrino  with emission of a photon $\nu \rightarrow \nu +\gamma$,  \cite{Neutrino-raddecay}
\item Photon decays into neutrino-antineutrino pair in plasma: $\gamma\rightarrow \nu \bar{\nu}$, \citep{Neutrino-Plasmon}
\end{itemize}
 Neutrino– one-photon interactions are of interest since they may play a key role in elucidating the solar neutrino puzzle, which can be explained by a large neutrino magnetic moment \cite{solar-neutrino} or a resonant spin flip induced by Majorana neutrinos \cite{Resonant-flip}.

A Dirac neutrino with standard model  (SM)  interactions has a magnetic moment which is given by \cite{numu-SM}
\begin{equation}
\mu_{\nu_i}=\frac{3m_eG_F}{4\sqrt{2}\pi^2 }m_{\nu_i} \approx 3\times 10^{-19}\left( \frac{m_{\nu_i}}{eV}\right) \mu_B
\end{equation}
where $\mu_B=\frac{e}{2m_e}$ is the Bohr Magneton. Current limits on these magnetic moments are several orders of magnitude
larger \cite{numu-Ex} therefore a magnetic moment close to these limits would
indicate a window for probing effects induced by new physics
beyond the SM \cite{Fukugita}. Similarly, a neutrino electric
dipole moment will point also to new physics and they will be of
relevance in astrophysics and cosmology, as well as terrestrial
neutrino experiments \cite{Cisneros}.

The current best limit on $\mu_{\nu_\tau}$ has been obtained  in the Borexino
experiment which explores solar neutrinos \cite{Borexino}. Some experimental limits on the magnetic moment of the tau-neutrino
are shown in Table I.

\begin{table}[ht]
\caption{Experimental limits on the magnetic moment of the tau-neutrino}
\begin{center}
\begin{tabular}{c| c| c| c| c}
\hline
Experiment     &  Method          &  Limit                                      & C. L.  &  Reference\\
\hline
\hline

Borexino       &  Solar neutrino  &  $\mu_{\nu_\tau} < 1.9\times 10^{-10}\mu_B$ & $90 \%$  & \cite{Borexino} \\
\hline
E872 (DONUT)   &  Accelerator $\nu_\tau e^-, \bar\nu_\tau e^-$ & $\mu_{\nu_\tau} < 3.9\times 10^{-7}\mu_B$  & $90 \%$  & \cite{DONUT} \\
\hline
CERN-WA-066    &  Accelerator     & $\mu_{\nu_\tau} < 5.4\times 10^{-7}\mu_B$   & $90 \%$ & \cite{WA66} \\
\hline
L3             & Accelerator      & $\mu_{\nu_\tau} < 3.3\times 10^{-6}\mu_B$   & $90 \%$  & \cite{L3} \\
\hline
\end{tabular}
\end{center}
\end{table}

The bound on $\mu_{\nu_\tau}$    was obtained through the analysis of the
process $e^{+}\, e^{-} \rightarrow \nu \, \bar \nu \, \gamma$   near the
$Z$-resonance, with a massive neutrino and the SM
$Z e^{+}\, e^{-}$ and $Z\nu\,  \bar \nu$  couplings \cite{T.M.Gould}. At low center of mass energy $s \ll  M^2_{Z}$,
 the dominant
contribution to the process $e^+\, e^- \to \nu \, \bar\nu\,  \gamma$
involves the exchange of a virtual photon \cite{H.Grotch}. The
dependence on the magnetic moment comes from a direct coupling to
the virtual photon, and the observed photon is a result of
initial-state Bremsstrahlung.

At higher scale near the $Z$ pole ($s\approx M^2_{Z}$,) the
dominant contribution involves the exchange of the $Z$ boson. The
dependence on the magnetic moment $(\mu_{\nu_\tau})$ and the
electric dipole moment $(d_{\nu_\tau})$ now comes from the
radiation of the photon observed by the neutrino or antineutrino
in the final state. We emphasize here the importance of the final
state radiation near the $Z$ pole of a very energetic photon as
compared to conventional Bremsstrahlung.

Additional neutral  gauge bosons appear in  most extended models of the SM such as Left-Right Symmetric Models (LRSM)
\cite{G.Senjanovic,G.Senjanovic1},
models of composite gauge bosons \cite{Baur} or the  $SU(3)_C \times SU(3)_L\times U(1)_X$ (3-3-1) models \cite{331}. In
particular, it is possible to study some phenomenological features
associated with this extra neutral gauge boson through models with
gauge symmetry $SU(3)_C\times SU(4)_L\times U(1)_X$, also called
3-4-1 models \cite{341}. In this model there exit two new neutral  gauge bosons which result in large constraint to the neutrino dipole moment.
In the framework of the 3-4-1 model, the puzzle of the large magnetic moment of neutrino with its small mass was firstly considered in Ref. \cite{o341}.


Let us mention with the current situation of the experimental bounds. The L3 collaboration \cite{L3} uses detector-simulated $e^+ \, e^- \rightarrow \bar{\nu\nu \gamma(\gamma)}$ events, random trigger events, and large angle  $e^+ e^- \rightarrow e^+ e^- $ events to evaluate the selection efficiency. In
Fig. 2   of \cite{L3} only 6 events was found as real background with the angular interval $-0.7<\cos{\theta_\gamma}<0.7$.
The event number  $N$ can be approximated as $N=N_B +\sqrt{N_B} $ where $N_B$ is the estimated background event and sufficient large ($N_B>10$). This means that limits on parameters at different confidential levels can be found by replacing the equation for the total number of expected events $N=N_B +\sqrt{N_B} $ in the expression $N\approx \sigma(\mu_{\nu_\tau}, d_{{\nu_\tau}} ) \cal{L} $  \cite{Nevents}.

 As discussed in \cite{L3,T.M.Gould} the total number of event was calculated  at $1\sigma $, $2\sigma $, $3\sigma $. Taking into account with the luminosity $\cal{L}$=500\, fb$^{-1}$ \cite{CLIC,ILC} we can obtain the limit of the neutrino magnetic moment and the neutrino electric dipole moment.

Our aim in this paper is to get bound of the  magnetic and   electric dipole moments of the neutrino by analyzing the reaction
$e^{+}e^{-}\rightarrow \nu \bar\nu \gamma$ in the framework of  the
$SU(4)_L \times U(1)_X$ model. We
will focus on  the anomalous magnetic moment (MM) and
the electric dipole moment (EDM) of  massive tau-neutrino. We will then
set limits on the tau-neutrino MM and EDM according to the ratio of the $SU(4)\times U(1)_X$ scale versus $SU(2)_L\times U(1)_Y$ scale.
Since the $W$ and photon exchange diagrams  amounting
 to just $1\hspace{.5mm} \%$ corrections  in the relevant kinematic regime,  will be neglected. To justify this argument, the reader is referred  to Ref.\cite{longamm}.
The Feynman diagrams which give the most important
contribution to the cross section are shown in Fig.\ref{Feynman-diagrams}. We will set limits on tau neutrino dipole moment for integrated luminosity of
500-2000\, fb$^{-1}$ and center of mass energy between 0.5 and 3.0 TeV which can be archive in the next generation of linear colliders, namely, the International Linear Collider (ILC)\cite{ILC} and the Compact Linear Collider (CLIC)\cite{CLIC}.

This paper is organized as follows: In Sec. \ref{sec2} we will briefly review the 3-4-1 model then in Sec. \ref{sec3} we present the
calculation of the process $e^{+}e^{-}\rightarrow \nu \bar\nu
\gamma$ in the context of a $SU(4)_L\times U(1)_X$ model. Finally, we present our
results and conclusions in Sec. \ref{sec4}.

\vspace{3mm}

\section{Minimal 3-4-1 with right-handed neutrinos}
\label{sec2}

The $SU(4)_L\times U(1)_X$ model was originally proposed in \cite{341-original}.
The minimal $SU(4)_L\times U(1)_X$ model was systematically  studied in \cite{341}. In this section we will quickly review the model.
The leptonic structure of the $SU(4)_L \times U(1)_X$ model is arranged as:
\begin{equation}
f_{aL}=(\nu_a, l_a,l^c_a,\nu^c_a)^T_L \sim (1,4,0) \hspace*{1cm} a=e, \mu, \tau
\end{equation}
where   $\nu^c_L\equiv (\nu_R)^c$ and the charge conjugation of
$f_{aL}$
\begin{equation}
f^c_{aR} \equiv (f_{aL})^c=(\nu^c_{aR},l^c_{aR},l_{aR},\nu_{aR} )^T
\end{equation}

One quark generation is arranged into quadruplet:
\begin{eqnarray}
Q_{3L}&=&(u_3,d_3,T,T')^T_L\sim \left(3,4,\frac{2}{3} \right) \nonumber  \\
u_{3R}&\sim&(3,1,2/3),  d_{3R}\sim (3,1,-1/3),   \nonumber \\
T_R &\sim & \ (3,1,5/3), T'_R  \sim (3,1,2/3).
\end{eqnarray}

The two other quark generations are arranged as antiquadruplet
\begin{eqnarray}
Q_{\alpha L}&=&(d_\alpha ,-u_\alpha ,D_\alpha ,D'_\alpha )^T_L\sim \left(3,4^*,-\frac{1}{3} \right), \alpha=1,2 \crn
u_{\alpha R}&\sim&(3,1,2/3),  d_{\alpha R}\sim (3,1,-1/3),   \crn
D_{\alpha R} &\sim & \ (3,1,-4/3), D'_{\alpha R}  \sim (3,1,-1/3).
\nn
\end{eqnarray}

The Higgs sector consists four Higgs quadruplets  given below
\begin{eqnarray}
\chi &=& (\chi^0_1, \chi^-_2, \chi^+_3,\chi^0_4  )^T\sim (1,4,0), \vspace*{1cm} \Phi=(\Phi^-_1, \Phi^{--}, \Phi^0,\Phi^-_2  )^T\sim (1,4,-1) \crn
 \rho &=& (\rho_1^+, \rho^0, \rho^{++},\rho_2^+)^T \sim (1,4,1), \vspace*{1cm} \eta=(\eta_1^0,\eta_2^-,\eta_3^+,\eta_4^0)^T \sim (1,4,0)\crn
\end{eqnarray}
and one symmetric decuplet (${\bf 10}_S$) as
\be  H \sim (1, {\bf 10}, 0) = \frac{1}{\sqrt{2}}\left(%
\begin{array}{cccc}
\sqrt{2} H_1^0 &H_1^-  & H_2^+& H_2^0 \\
 H_1^- & \sqrt{2} H_1^{--}     & H_3^0 &  H_3^- \\
H_2^+ & H_3^0  &\sqrt{2} H_2^{++}&H_4^+ \\
H_2^0 &H_3^-&H_4^+&\sqrt{2} H_4^0\
\end{array}\,
\right)\, . \label{Higg62}
\ee
The necessary  vacuum  expectation value  (VEV)  structure is given by
\bea \langle \chi \rangle & =  &  \left(%
0 \, ,
0\, ,
0\, ,
\fr{V}{\sqrt{2}}
\right)^T \, , \hs
 \langle \ph \rangle  =    \left(%
0 \, ,
0\, ,
\fr{\om}{\sqrt{2}}\, ,
0
\right)^T ,\crn
\langle \rho \rangle & =  &  \left(%
0\, ,\fr{v}{\sqrt{2}}\, ,
0\, , 0
\right)^T \, , \hs
\langle \eta \rangle  =    \left(%
\fr{u}{\sqrt{2}}\, ,
0\, , 0\, ,
0
\right)^T,
\label{l63}
\eea
and
 \be  \langle H \rangle  =\fr{1}{2}\left(%
\begin{array}{cccc}
0 & 0 &  0 & \epsilon \\
 0 & 0   & v' &  0 \\
0 & v'  & 0 &0 \\
\epsilon &0 &0&0\
\end{array}\,
\right)\, . \label{Higg14}
\ee
Then all fermions and gauge bosons get necessary masses \cite{341}.

  In the model,  the gauge sector consists six charged/non-Hermitian gauge bosons and four neutral ones. 
The charged and non-Hermitian neutral gauge bosons defined through
\bea P_\mu^{CC} & = &\fr 1 2 \sum_{a} \la_a A_{a} \, , \hs a = 1,2,4,5,6,7,9,10,11,12,13,14\crn
&=&\fr{1}{ \sqrt{2}}\left(%
\begin{array}{cccc}
0 & W^{'+} &W_{13}^{-} & W_{14}^{0}\\
 W^{'-}  & 0 & W_{23}^{--}  &W_{24}^{-}\\
W_{13}^{+} & W_{23}^{++} & 0 &W_{34}^{+} \\
\left(W_{14}^{0}\right)^*&W_{24}^{+}&W_{34}^{-}&0
\end{array}\,
\right)_\mu
=\fr{1}{ \sqrt{2}}\left(%
\begin{array}{cccc}
0 & W^{'+} &Y^{'-}& N^{0}\\
 W^{'-}  & 0 & U^{--} &X^{'-}  \\
Y^{'+} & U^{++} & 0 &K^{'+}\\
\left(N^{0}\right)^*&X^{'+}& K^{'-}&0
\end{array}\,
\right)_\mu
\label{m4eq258}
\eea

The above gauge bosons mix each other, and the physical states are determined as \cite{341}
\be W_\mu  =  \cos \theta \, W'_\mu - \sin \theta\, K'_\mu\, , \hs
K_\mu  =   \sin \theta\,  W'_\mu + \cos \theta\, K'_\mu\, ,
\label{l66}
 \ee
 where the mixing angle $\theta$ characterizing lepton number violation is given by
\be
\tan 2\theta  =   \fr{4v' \ep}{V^2 + \om^2 - u^2-v^2}\, .
\label{eq6}
\ee

For the $X-Y$ mixing, we obtain the physical states
\be Y_\mu  =  \cos \theta' \, Y'_\mu - \sin \theta'\, X'_\mu\, ,\hs
X_\mu   =   \sin \theta'\,  Y'_\mu + \cos \theta'\, X'_\mu\,
\label{l67}
 \ee
with  the mixing angle  defined as
\be
\tan 2\theta'  =   \fr{4v' \ep}{V^2 - \om^2 - u^2 +v^2}\, .
\label{eq7}
\ee
The masses of physical gauge bosons  are determined as
\bea m^2_{W^\pm} &\simeq & \fr{g^2}{4} (v^2+u^2+v''^2) \,  ,\hs m^2_{K^\pm}
 \simeq  \fr{g^2}{4} ( V^2+ w^2+v''^2) \,  ,\crn
m^2_{X^\pm} &\simeq & \fr{g^2}{4}(V^2+v^2+v''^2) \,  ,\hs m^2_{Y^\pm}
 \simeq  \fr{g^2}{4} (w^2+u^2+v''^2) \,  .
\label{lm82}
\eea

The four  neutral gauge bosons are the photon and three neutral gauge bosons labeled by $Z_i, i= 1, 2, 3$.

\subsection{Charged currents}

Taking into account of  the mixing among singly charged gauge bosons, we can express above expression as
follows
\bea - \mathcal{L}^{\mathrm{CC}}=\fr{g}{\sqrt{2}}\left(J^{\mu-}_W W^+_\mu + J^{\mu-}_K K^+_\mu + J^{\mu -}_X X^{+}_\mu +
J^{\mu-}_Y Y^+_\mu + J^{\mu 0*}_N N^{0}_\mu +  J^{\mu --}_U U^{++}_\mu +\mathrm{H.c.}\right), \eea
where
\bea J^{\mu-}_W&=&c_\theta (\overline{\nu}_{aL}\ga^\mu
l_{aL}+ \overline{u_{3 L}} \ga^\mu d_{3 L}- \overline{u}_{\al L}\ga^\mu d_{\al L})
\crn &-& s_\theta
(- \overline{ \nu_{a R}} \ga^\mu  l_{a R}   + \overline{T_L} \ga^\mu  T_L^\prime +
 \overline{D^\prime_{\al L}} \ga^\mu D_{\al L}),  \label{lm61}\\
 J^{\mu-}_K&=&c_\theta (- \overline{ \nu_{a R}} \ga^\mu  l_{a R}   + \overline{T_L} \ga^\mu  T_L^\prime +
 \overline{D^\prime_{\al L}} \ga^\mu D_{\al L}) + s_\theta
 (\overline{\nu}_{aL}\ga^\mu
l_{aL}+ \overline{u_{3 L}} \ga^\mu d_{3 L} - \overline{u}_{\al L}\ga^\mu d_{\al L}),\crn
J^{\mu -}_X &=& c_{\theta'}(\overline{\nu^c_{a L}} \ga^\mu l_{a L}  + \overline{ T'_L} \ga^\mu d_{3 L}
- \overline{u_{\al L}} \ga^\mu D_{\al L}^\prime) +
s_{\theta'}( \overline{l^c_{a L}} \ga^\mu  \nu_{a L} + \overline{T_{ L}} \ga^\mu u_{3 L} + \overline{d_{\al L} } \ga^\mu  D_{\al L})\, , \crn
J^{\mu -}_Y &=& c_{\theta'}
( \overline{l^c_{a L}} \ga^\mu  \nu_{a L} + \overline{T_{ L}} \ga^\mu u_{3 L} + \overline{d_{\al L} } \ga^\mu  D_{\al L})
- s_{\theta'}
(\overline{\nu^c_{a L}} \ga^\mu l_{a L}  + \overline{ T'_L} \ga^\mu d_{3 L}
- \overline{u_{\al L}} \ga^\mu D_{\al L}^\prime)\, ,
 \crn
J^{\mu --}_U & = & \overline{l^c_{a L}} \ga^\mu  l_{a L}  + \overline{T_L} \ga^\mu d_{3 L}
  - \overline{u_{\al L}} \ga^\mu  D_{\al L}\, ,
\crn
J^{\mu 0*}_N & = & \overline{\nu_{a L}} \ga^\mu \nu^c_{a L} +
\overline{u_{3 L}} \ga^\mu T^\prime_{ L} +
\overline{D_{\al L}^\prime } \ga^\mu
 d_{\al L}.
\label{lmt} \eea
For precision,
in the quark sector the CKM matrix will be appeared.
In terms of mass eigenstates, the current in (\ref{lm61})
has a new  form
\bea
J^{\mu-}_W&=&c_\theta (\overline{\nu}_{iL}\ga^\mu V^{i j}_{PMNS}
l_{jL}+  s_\theta
 \overline{ \nu_{i R}} \ga^\mu V^{i j}_{PMNS} l_{j R}) + \cdots
\label{lm61t}
\eea

\subsection{Neutral current}


The Lagrangian of the fermion is
\begin{equation}
L=i \sum_f \bar{f}\gamma^\mu D_\mu f  +H.c.
\end{equation}
The Lagrangian for neutral current extracted from above Lagrangian is:
\begin{equation}
L^{NC}=g\bar{f}\gamma^\mu P_\mu^{NC} f
\end{equation}
 where $P_\mu^{NC}$ is given in \citep{341}.
Explicitly, the neutral current including the electromagnetic current are:

\begin{equation}
-L^{NC}=e J^{\mu}_{em}A_\mu+ \frac{g}{2c_w}\sum_{i=1}^3Z^i_\mu\sum_f\bar{f}\gamma^\mu[g^i_V(f)-g^i_A(f)\gamma_5]f
\end{equation}
 where
 \begin{equation}
 e=g\sin\theta_W, t=\frac{g'}{g}=\frac{2\sqrt{2}\sin\theta_W}{\sqrt{1-4\sin^2\theta_W}}
 \end{equation}
 $Z^{1,2,3}$ can be identified as $Z^1 \approx Z$, and $Z^{2,3}\approx Z'_{3,4}$ are exact eigenstates.

From  explicit calculation,  the needed couplings are given by
\begin{eqnarray}
g^1_V(e)&=&\frac{c_W (-3 c_{32}s_W -\sqrt{3}c_W) }{2\sqrt{3}}, \hspace*{1cm} g^1_A(e)=\frac{c_W( c_{32}s_W -\sqrt{3}c_W)}{2\sqrt{3}},
 \crn
 g^1_V(\nu)&=&c_W\left(\frac{1}{2}c_W-\frac{c_{32}s_W}{2\sqrt{3}} \right), \hspace*{1cm} g^1_A(\nu)=c_W\left(\frac{1}{2}c_W-\frac{c_{32}s_W}{2\sqrt{3}} \right), \crn
g^2_V(e)&=&-\frac{c_W c_\alpha c_{32}}{2 \sqrt{3}}, \hspace*{1cm} g^2_A(e)=\frac{c_W(c_\alpha c_{32}+\sqrt{2} s_\alpha)}{2\sqrt{3}},
 \crn
 g^2_V(\nu)&=&-\frac{c_W(c_\alpha s_{32} - 2 s_\alpha)}{2\sqrt{3}}, \hspace*{1cm} g^2_A(\nu)=\frac{c_W(c_\alpha s_{32}
 +\sqrt{2}s_\alpha)}{2\sqrt{3}},\crn
g^3_V(e)&=&-\sqrt{3}c_W s_\alpha s_{32} , \hspace*{1cm} g^3_A(e)=\frac{c_W(c_\alpha + s_\alpha s_{32})}{2\sqrt{3}}, \crn
 g^3_V(\nu)&=&\frac{c_W(2\sqrt{2}c_\alpha - s_\alpha s_{32})}{2\sqrt{3}}, \hspace*{1cm} g^3_A(\nu)=\frac{c_W(\sqrt{2} c_\alpha
 +s_\alpha s_{32})}{2\sqrt{3}},\nn
\end{eqnarray}
 where $s_W \equiv \sin\theta_W, \, c_W \equiv \cos\theta_W$ and
\begin{eqnarray}
s_{32}&=&\frac{2\sqrt{2}}{\sqrt{8+3t^2}}, \hspace*{1cm} c_{32}=\frac{-\sqrt{3}t}{\sqrt{8+3t^2}} ,  \crn
	 t_{2\alpha}&=&\frac{2\sqrt{8+3t^2}w^2}{9V^2-(7+3t^2)w^2}
\end{eqnarray}

\section{The total cross section}
\label{sec3}

\begin{figure}
  \includegraphics[scale=0.9]{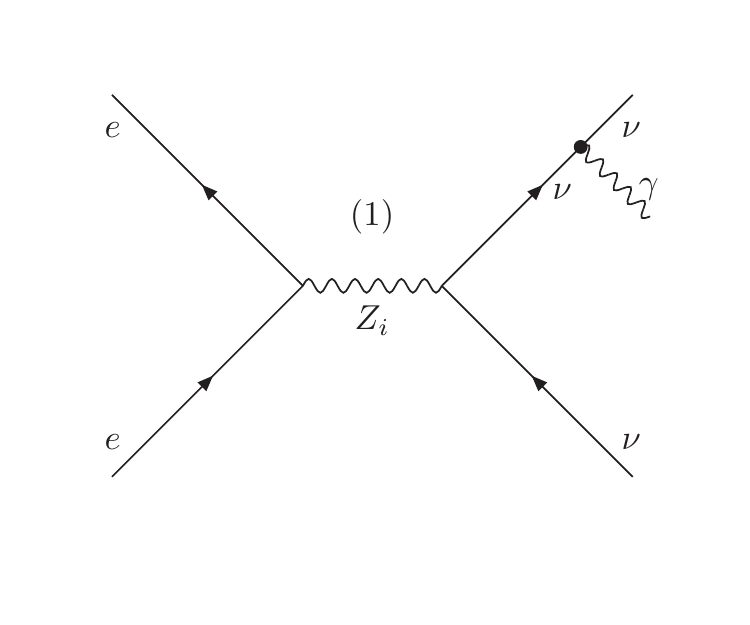}
    \includegraphics[scale=0.9]{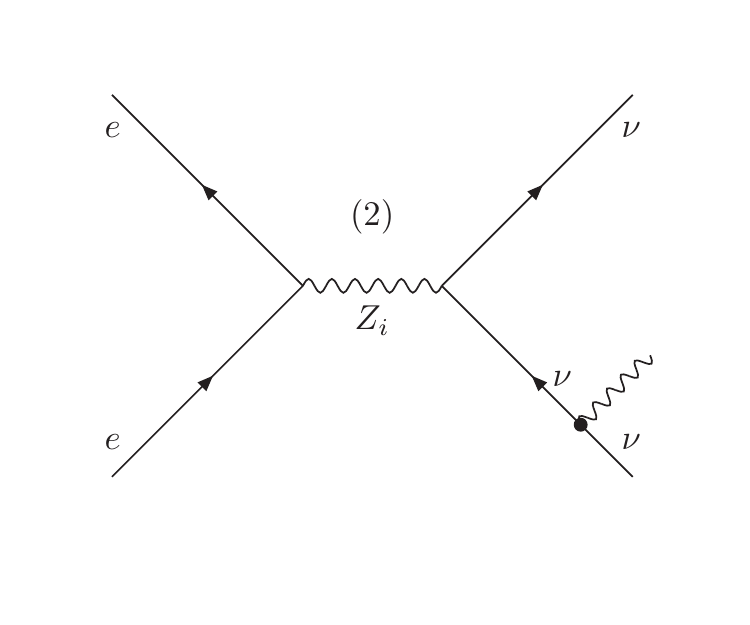}

  \caption{Feynman diagrams contribution to process $e^+e^- \rightarrow \bar{\nu} \nu \gamma$ in the $SU(4)_L \times U(1)_X$ model}
  \label{Feynman-diagrams}
\end{figure}

\vspace{3mm}%
 The total cross section of the process $ e^+ e^- \rightarrow \bar{\nu}_\tau \nu_\tau \gamma $ can be calculated using Breit-Wigner resonance form \cite{PDG}:

\begin{equation}
\sigma(e^+ e^- \rightarrow \bar{\nu}_\tau \nu_\tau \gamma )=\sum_{i=1}^{3}\frac{4\pi(2J+1)\Gamma_{e^+e^-}\Gamma_{\bar{\nu}_\tau \nu_\tau \gamma }}{(s-M^2_{Z_i})^2+M^2_{Z_i}\Gamma^2_{Z_i}}
\label{crosssection}
\end{equation}
where $Z_i, i=1,2,3$ are the SM Z boson and two new neutral bosons respectively and $\Gamma_{e^+e^-}$,$\Gamma_{\bar{\nu}_\tau \nu_\tau \gamma }$ are the respectively decay width of $Z_i$ in the channel $e^+e^-$ and $\nu \bar{\nu},\gamma$ (see Figure \ref{Feynman-diagrams}).


The decay of $Z_i$ to $e^+ e^-$ has the same structure as the decay of Z boson in to $e^+ e^-$.
The decay of Z boson in to $e^+ e^-$ is given by \cite{PDG}.
\begin{equation}
\Gamma_{(Z_1\rightarrow e^+ e^-)}=\frac{\alpha M_{Z_i}}{12}\frac{[g_V^2(e)+g_A^2(e)]}{s^2_W(1-x_2)}
\end{equation}
where
$\alpha=\frac{e^2}{4\pi}$ is the fine structure constant.

The decay rate of $Z_i$ to $e^+ e^-$ can be calculated as:
\begin{equation}
\Gamma_{(Z_i\rightarrow e^+ e^-)}= \Gamma_{(Z_1\rightarrow e^+ e^-)}\frac{M_{Z_i}}{M_{Z_1}}\frac{[g^2_{iV}(e)+g^2_{iA}(e)]}{[g^2_{1V}(e)+g^2_{1A}(e)]}\, .
\end{equation}

 In the followings we will investigate the decay of
$Z_i \rightarrow \bar{\nu} \nu \gamma$. The Feynman diagrams of this decay is given in Fig.\ref{Feynman-diagrams}
\begin{eqnarray}
{\cal M}_{1}(\bar{\nu} \nu \gamma)&&= \epsilon^{\mu}_{\delta}(p) \epsilon^{\beta}_{\lambda}(q) \left[ \bar
u(p_{3})\Gamma^{\mu} \frac{i(k\!\!\!/+m_{\nu})}{(k^{2}-m^{2}_{\nu})}\frac{(-ig)}{2c_W}\gamma^{\beta}
[g_{iV}(\nu)-g_{iA}(\nu)\gamma_5]v(p_{4})\right]
\end{eqnarray}

\begin{eqnarray}
{\cal M}_{2}(\bar{\nu} \nu \gamma)&&= \epsilon^{\mu}_{\delta}(p) \epsilon^{\beta}_{\lambda}(q) \left[ \bar
u(p_{3})\frac{(-ig)}{2c_W}\gamma^{\beta}
[g_{iV}(\nu)-g_{iA}(\nu)\gamma_5] \frac{i(k\!\!\!/+m_{\nu})}{(k^{2}-m^{2}_{\nu})} \Gamma^{\mu}  v(p_{4})\right]\, ,
\end{eqnarray}
where

\begin{equation}
\Gamma^{\alpha}=eF_{1}(q^{2})\gamma^{\alpha}+\frac{ie}{2m_{\nu}}F_{2}(q^{2})\sigma^{\alpha
\mu}q_{\mu}+eF_3(q^2)\gamma_5\sigma^{\alpha\mu}q_\mu,\nn
\end{equation}
 is the tau-neutrino electromagnetic vertex, $e$ is the
charge of the electron, $q^\mu$ is the photon momentum and
$F_{1,2,3}(q^2)$ are the electromagnetic form factors of the
neutrino, corresponding to charge radius, MM and EDM,
respectively, at $q^2 = 0$ \cite{R. Escribano},
while $\epsilon^\lambda_\alpha$ is the polarization vector of the
photon. $p$ and $k$ stand for the momenta  of the Z and neutrino, respectively.

Summing over spin, the square of the scattering amplitude is:
\begin{equation}
\sum_s\mid{\cal M}_{1,2}\mid^2=\frac{g^2}{4\cos^2\theta_W}(\mu^2_{\nu_\tau}+d^2_{\nu_\tau})
[( g^2_{iV}(\nu)+ g^2_{iA}(\nu))(s-2\sqrt{s}E_\gamma)+g^2_{iA}(\nu)E^2_\gamma\sin^2\theta_\gamma],
\end{equation}

The decay rate of gauge boson $Z_i \rightarrow \nu \bar\nu \gamma$ is therefore calculated as:
\begin{eqnarray}
\Gamma_{(Z_i \rightarrow \nu \bar\nu \gamma)}&=&\int\frac{\alpha(\mu_{\nu_\tau}^2+d_{\nu_\tau}^2)}{64\pi^2M_{Z_{i}}x_W(1-x_W)}[( g^2_{iV}(\nu)+ g^2_{iA}(\nu))(s-2\sqrt{s}E_\gamma)+g^2_{iA}(\nu)E^2_\gamma\sin^2\theta_\gamma]  \nonumber \\
&& E_\gamma dE_\gamma d\cos\theta_\gamma\, .
\end{eqnarray}

Substituting above expression into (\ref{crosssection}) we have the total cross section of the process $ e^+ e^- \rightarrow \bar{\nu}_\tau \nu_\tau \gamma $:

\begin{eqnarray}
\sigma(e^{+}e^{-}\rightarrow \nu \bar\nu \gamma)&=&\sum_{i=1,2,3} \int E_\gamma dE_\gamma d\cos\theta_\gamma\frac{\alpha^2(\mu_{\nu_\tau}^2+d_{\nu_\tau}^2)}{192\pi}
\frac{ [g^2_{iV}(e)+ g^2_{iA}(e)]}{x^2_W(1-x_W)^2}
\crn
&& \times \frac{[( g^2_{iV}(\nu)+ g^2_{iA}(\nu))(s-2\sqrt{s}E_\gamma)+g^2_{iA}(\nu)E^2_\gamma\sin^2\theta_\gamma] }{(s-M^2_{Z_{i}})^2+M^2_{Z_{i}}\Gamma^2_{Z_{i}}}\, .
\end{eqnarray}

\begin{figure}
    \includegraphics[scale=0.5]{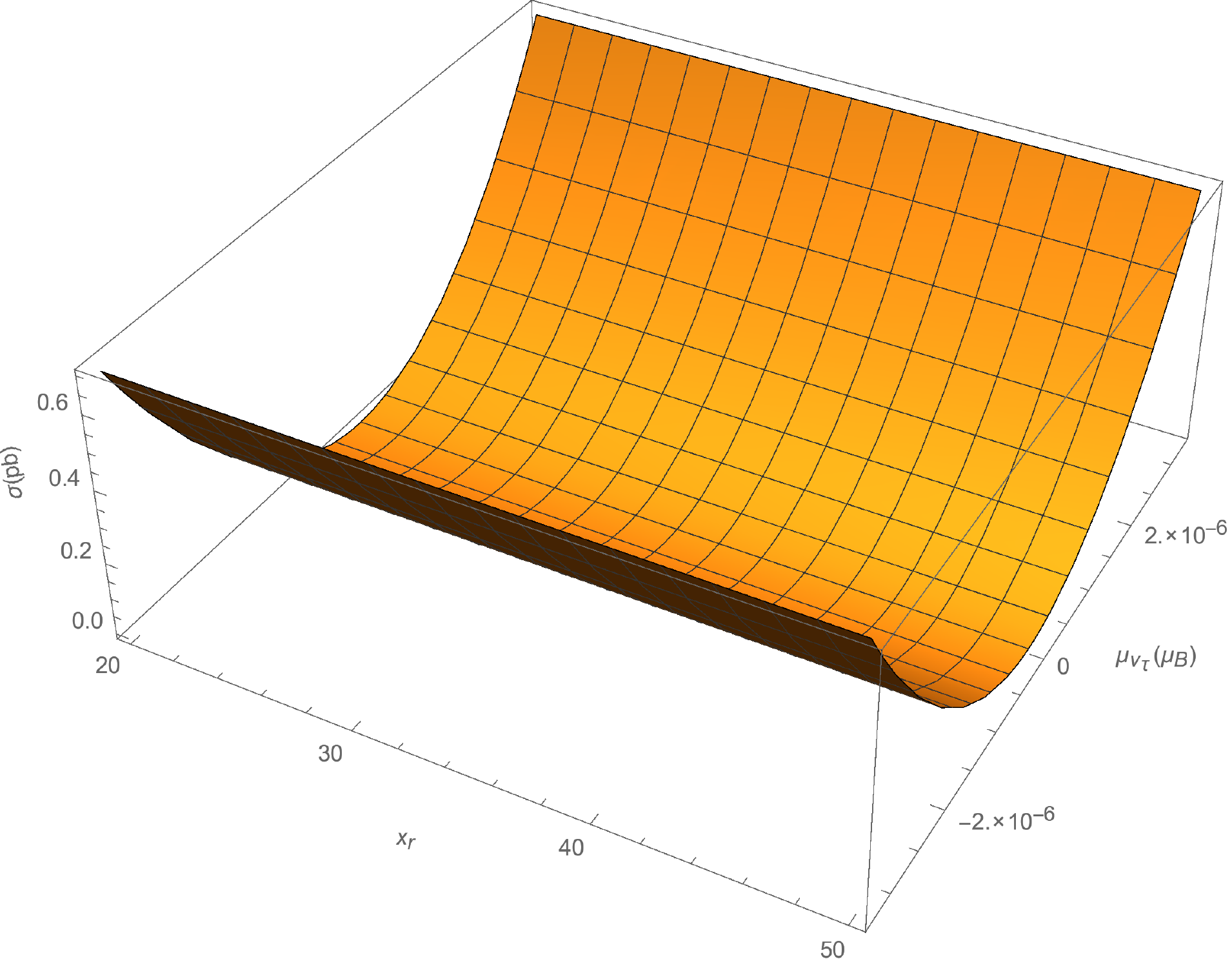}

  \caption{The total cross section for  $e^+e^- \rightarrow \bar{\nu} \nu \gamma$ as the function of the ration of $\frac{m_{Z'}}{m_{Z}}$ and $\mu_{\nu_\tau}$}
  \label{cross}
\end{figure}

\section{Results}
\label{sec4}

In investigating numerically the  cross section, the photon angle and energy will be cut to avoid the divergence of the integral when integrating over important intervals.  We integrate over $\theta_\gamma$ from $44.5^o$ to $135.5^o$ and $E_\gamma$ from 15 GeV to 100 GeV. The following numerical values are used: sin$^2\theta_W=0.23126 \pm 0.00022 $, $M_Z=91.1876 \pm 0.0021 GeV$, $\Gamma_Z=2.4952 \pm 0.0023 GeV$.   We approximate  the mass of the two new neutral bosons are of the same order
$(M_{Z_2}\approx M_{Z_3})$ therefore their decay rate can be approximate to have the same order $\Gamma_{Z_2} \approx \Gamma_{Z_3}$.
The decay width of the $Z_2$, $Z_3$ bosons are approximate as: $\Gamma_{Z_{2,3}}=2 \Gamma_{Z_1}$ \cite{Z23decayrate} and the mass of $Z_2$, $Z_3$ bosons can be approximate as $M_{Z_{2,3}}=x_r M_{Z_1}$ where $x_r=\frac{M_{Z_2}}{M_{Z_1}}$. The mass limit of the new neural gauge boson is $M_{Z'}<5$ TeV \cite{Z'masslimit} equivalent to $x_r\approx 50$. We obtain the total cross section $\sigma_{Tot}=\sigma_{Tot}(\mu_{\nu_\tau},d_{\nu_\tau},\sqrt{S},x_r)$. We will evaluate the total cross section as a function of the parameters of the model, $x_r$, which is the ratio of the symmetry breaking scale of the group $SU(4)_L$ and the vacuum expectation value of the $SU(2)_L$ and the center of mass energy.    Using the approximation  that the total number events  N $ \approx \sigma_{Tot} \cal{L}$ \cite{Nevents} where $N=N_B+\sqrt{N_B}$ B is the total number of $e^+e^− \rightarrow \bar{\nu_\tau}\nu_\tau\gamma $ events expected at $1\sigma, 2 \sigma, 3 \sigma$ we can set the bounds of the tau neutrino magnetic dipole moments with $d_{\nu_\tau}=0$ for different integrated luminosity $\cal{L}$. This analysis can be used to obtain the bound on the tau neutrino electric dipole moment with $\nu_\tau=0$

\begin{table}[!ht]
\begin{center}
\begin{tabular}{ccc}
  \hline
  \multicolumn{3}{c}{${\cal L}=500,\hspace{0.8mm} 1000,\hspace{0.8mm} 2000\hspace{0.8mm}fb^{-1}$}\\
   \hline

  \multicolumn{3}{c}{$ \sqrt{S} =1.0 \hspace{0.8mm}TeV; x_r=20$ }\\

  C.L. & $| \mu_{\nu_\tau}(\mu_B) |$ & $| d_{\nu_\tau}(e.cm) |$\\
  \hline
  1$\sigma$ & $(0.35, 0.25, 0.17)\times 10^{-10}$ & $(1.03, 0.73, 0.51)\times 10^{-19}$\\
  2$\sigma$ & $(0.40, 0.28, 0.20)\times 10^{-10}$ & $(1.18, 0.84, 0.59)\times 10^{-19}$\\
  3$\sigma$ &$(0.41, 0.29, 0.20)\times 10^{-10}$ & $(1.22, 0.86, 0.61)\times 10^{-19}$\\
  \hline
   \multicolumn{3}{c}{$ \sqrt{S} =2.0 \hspace{0.8mm}TeV; x_r=35$ }\\

  C.L. & $| \mu_{\nu_\tau}(\mu_B) |$ & $| d_{\nu_\tau}(e.cm) |$\\
  \hline
  1$\sigma$ & $(0.51, 0.36, 0.25)\times 10^{-10}$ & $(2.0, 1.43, 1.01)\times 10^{-19}$\\
  2$\sigma$ & $(0.58, 0.41, 0.29)\times 10^{-10}$ & $(2.34, 1.65, 1.17)\times 10^{-19}$\\
  3$\sigma$ &$(0.60, 0.42, 0.30)\times 10^{-10}$ & $(2.4, 1.70, 1.20)\times 10^{-19}$\\
  \hline
    \multicolumn{3}{c}{$ \sqrt{S} =3.0\hspace{0.8mm} TeV; x_r=50$ }\\

  C.L. & $|\mu_{\nu_\tau}(\mu_B)|$ & $|d_{\nu_\tau}(e.cm)|$\\
  \hline
  1$\sigma$ & $(1.02, 0.72, 0.51)\times 10^{-10}$ & $(3.0, 2.12, 1.50)\times 10^{-19}$\\
  2$\sigma$ & $(1.18, 0.83, 0.59)\times 10^{-10}$ & $(3.46, 2.49, 1.73)\times 10^{-19}$\\
  3$\sigma$ &$(1.21, 0.85, 0.60)\times 10^{-10}$ & $(3.55, 2.51, 1.78)\times 10^{-19}$\\
  \hline
  \hline
\end{tabular}
		\caption{Bounds on the $\mu_{\nu_\tau}$ magnetic moment and $d_{\nu_\tau}$ electric  dipole moment for $\sqrt{S}=1, 2, 3 $ TeV and
$\cal{L}$=500, 1000, 2000 fb$^{-1}$ at 1$\sigma$, 2$\sigma$, 3$\sigma$ }
\end{center}
\end{table}

We present the
bounds obtained on the $\mu_{\nu_\tau}$ magnetic moment and $d_{\nu_\tau}$
electric dipole moment in Table II to demonstrate the order of magnitude.  From Table II we can see that our result is better than those given in literature \cite{T.M.Gould,CLIC,M. Acciarri,R. Escribano,tauneu-mmboud1,tauneu-mmboud2,tauneu-mmboud3,tauneu-mmboud4}
and approach the limit by Borexino experiment \cite{Borexino}.

In the case of the electric dipole moment  our result show that these bounds
are of order $10^{-19}$-$10^{-20}$ for the 95\% C.L. sensitivity limits at 1000 -
3000 GeV center of mass energies and integrated luminosities of 2000 $f b^{−1}$.
These bounds are improved by 2-3 orders of magnitude than those reported in the literature
\cite{tauneu-mmboud1,tauneu-mmboud2,tauneu-mmboud5,tauneu-mmboud6}

In Fig. \ref{cross} we evaluate the cross-section of the process   $e^+e^- \rightarrow \bar{\nu} \nu \gamma$ as the function of the ration of $\frac{m_{Z'}}{m_{Z}}$ and $\mu_{\nu_\tau}$ with the value of $\sqrt{S}=500$GeV. The value of the magnetic moment investigated is up to the current value of the L3 experiment $\mu_{\nu_\tau}=3.3\times 10^{-6}(\mu_B)$. Our result is of the same order with previous result \cite{tauneu-mmboud5}

In summary, we conclude that the estimated limits for the tau-neutrino magnetic and
electric dipole moments in the context of a $SU(4)_L \times U(1)_X$ model compare favorably with the limits
obtained by the L3 Collaboration and complement previous studies on the dipole moments.

\section*{Acknowledgments}

This research is funded by the Vietnam National Foundation for Science and Technology
Development (NAFOSTED) under grant number 103.01-2017.356


\end{document}